\documentstyle[epsf,a4wide]{article}

\begin{document} 
\openup1\jot
\begin{tabbing}
\hskip 14 cm \= {MRAO-1900}\\
\> DAMTP-96-11\\
\> Submitted to {\em PRL}\\
\> February 1996 \\
\end{tabbing}

\begin{center}
{\Huge\bf  
Cosmic Microwave Background experiments targeting the cosmic strings
Doppler peak signal  }
\vskip 1.2cm
{ Joao Magueijo$^{1,2}$, Mike P. Hobson$^2$ }
\vskip 1.2cm
$^{(1)}$Department of Applied Mathematics and Theoretical
Physics, University of Cambridge, Cambridge CB3 9EW, U.K.\\
$^{(2)}$Mullard Radio Astronomy Observatory,
Cavendish Laboratory, Madingley Road, Cambridge, CB3 0HE, U.K.
\end{center}

\begin{abstract}
\openup1\jot
We investigate which experiments are better suited to test the 
robust prediction that cosmic strings do not produce secondary Doppler
peaks.  We propose a statistic for detecting  oscillations in the 
$C^l$ spectrum, and study its statistical relevance given the truth
of an inflationary competitor to cosmic strings. The analysis is performed
for single-dish experiments 
and interferometers, subject to a variety of noise
levels and scanning features. A high resolution of 0.2 degrees
is found to be required for single-dish experiments 
with realistic levels of noise.
Interferometers appear to be more suitable for detecting this signal.
\end{abstract}

\setcounter{footnote}{0}
\newpage
In two recent {\it Letters} \cite{us,us2} 
it was shown how the existence or absence of
secondary Doppler peaks in the cosmic microwave background (CMB)
power spectrum $C^l$ could rule out or confirm 
a large class of defect theories, including cosmic strings.
The argument used is  attractive in that it does not depend on details
of defects or inflation, but only makes use of well-established
contrasting properties peculiar to the two types of theory. 
This is particularly welcome when calculations in defect scenarios
are so difficult and unreliable. A simple but robust test based on an issue
about to be decided by experiment
seems a soundly cautious approach to defect theories.

The question therefore arises: which CMB experiments can resolve the
secondary Doppler peaks? This is a timely issue
when so many proposals for ground based and satellite borne CMB
experiments are being made \cite{exp}. Experimental features have so far
been motivated by their implications on  inflationary parameter fixing 
from  Doppler
peaks' position, height, and  shape \cite{knox,jung}. 
Secondary peak detection is a far less demanding task, 
and can be used to quantify experimental spectral
resolution at the most basic level. In \cite{mikeandjoao} we analyze this
problem in a general setting, but here we consider only its bearings on
cosmic string theories. 

\begin{figure}
\begin{center}
    \leavevmode
    {\hbox %
   {\epsfxsize = 8cm \epsfysize = 8cm
    \epsffile {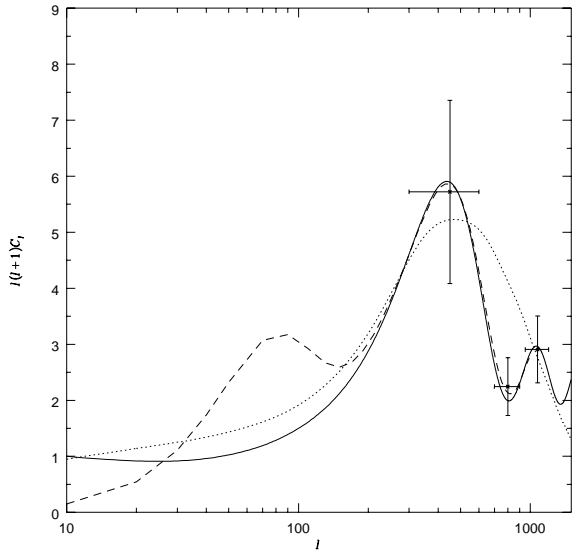} }}
\end{center}
\caption{The angular power spectrum of stCDM (line) and one possibility
for cosmic strings (points). The dashed curve is the stCDM power spectrum 
$C^s(k)$ as sampled by an interferometer with $\theta_w=2^o$
(divided by $\alpha$ defined in Eqn.~(\ref{cscn})). 
We have plotted the bins used in ${\cal C}$ as horizontal bars, their
level indicating the average spectrum in bin. The vertical
errorbars are the cosmic/sample variance.}
\label{fig1}
\end{figure}

Apart from the absence of secondary Doppler peaks in cosmic string
theories, the only reliable feature known is that 
their primary peak is located 
at  $l=400-600$. If the main peak is measured outside this range one
can rule out cosmic strings, but we shall assume here that this is not
the case. In order to identify experiments targeting the string's
lack of secondary oscillations we investigate how we could
falsify cosmic strings given the truth of a competitor inflationary
scenario with a primary peak located in $l=400-600$ and exhibiting 
secondary oscillations. For definiteness
we have taken CDM with $\Omega=0.3$, $h=0.6$, $\Omega_b h^2=0.02$,
and a flat primordial spectrum. We shall call this theory 
stCDM, the CDM competitor of cosmic strings, and we plot its $C^l$ spectrum
in Fig~\ref{fig1}. The main peak height
and shape will be assumed to be the same for cosmic strings
and stCDM, and the low $l$ section of the spectrum will be ignored.
In this way we assume maximal confusion in whatever is uncertain,
or alien to the signal to be experimentally tested. 

The idea is to apply to stCDM a statistic sensitive only to the
existence or absence of secondary oscillations. 
To set it up, we first compute the average power
$C_i$ in bins $i=1,2,3$ denoted by  horizontal bars
in Fig~\ref{fig1}. We then infer the spectrum convexity with
${\cal C}=(C_{1}+C_{3})/2 -C_2$.  A positive convexity reflects
unambiguously the first dip of the spectrum and therefore the first secondary 
oscillation. If the overall error in ${\cal C}$ is 
$\sigma({\cal C})$  then one can claim that ${\cal C}$ is
positive (and therefore that there are secondary oscillations) with
a number of sigmas equal to 
\begin{equation}
{\Sigma}={{\langle {\cal C}\rangle}\over \sigma
({\cal C})}\; .
\end{equation}
$\Sigma$ is then
to be seen as the stCDM secondary peak detection function,
or equivalently, the cosmic string rejection function.

In this {\it Letter} we set up a large parameter space of experiments,
on which we compute the contours of $\Sigma$.
We consider two types of experiments: single-dish experiments
(recovering some of the results in \cite{jung})
and interferometers. For single-dish experiments we allow 
the beam size, sky coverage, and detector noise to vary. For
interferometers we take as free parameters
the primary beam, number of fields,
and detector noise. We consider errors associated with
cosmic/sample variance, spectral resolution limitations 
due to finite sky coverage, and instrumental noise.
Foreground subtraction errors are included (not naively, as we prove 
in \cite{mikeandjoao}) in the form of only an extra instrumental noise term. 

We outline a method for computing errors in $C^l$ estimates
explained in more detail and generality in \cite{mikeandjoao}. 
For simplicity we consider the small field limit, and assume at first no
instrumental noise or foreground subtraction uncertainties.
We stereographically project the sky onto a plane, 
and expand in Fourier modes, using
the symmetric notation (factors of $2\pi$ evenly distributed). 
We denote by $a({\bf k})$ the modes provided
by an all-sky observation with infinite resolution, and $a^s({\bf k})$ the
modes as seen through an observation window $W({\bf x})$ and
convolved with a beam $B({\bf x})$.
For a single-dish experiment we shall assume that the window is a square
of side $L$ treated with a cosine bell \cite{teggy} 
(to bar edge effects),
and that the beam is a Gaussian with FWHM $\theta_b$. For interferometers
the window (better known as the primary beam) is a Gaussian with FWHM
$\theta_w$, and the beam is essentially unity \cite{mike}.
Using the convolution theorem twice we have that 
\begin{equation}\label{as}
a^s({\bf k})={\int {d^2k'\over 2\pi} a({\bf k'})B({\bf k'})
W({\bf k}-{\bf k'})}  
\end{equation}
where $W({\bf k})$ is the window Fourier transform.
The all-sky modes form a diagonal covariance matrix
${\langle a({\bf k})a^{\star}({\bf k'}) \rangle}
=C(k)\delta({\bf k}-{\bf k'})$,
where the brackets denote ensemble averages. 
In calculations concerning small patches 
of the sky $C(k)$ can be obtained 
by interpolating the $C^l$ 
with $k=l$ \cite{mikeandjoao,be}. 
On the other hand the sampled modes 
$a^s({\bf k})$ form the covariance matrix \cite{mike}
\begin{eqnarray}\label{covmat}
&{\langle a^s({\bf k})a^{s*}({\bf k'})\rangle}=&\nonumber\\
&{\int {d{\bf k}^{''}\over (2\pi)^2} C({\bf k^{''}})
B^2({\bf k^{''}})W({\bf k}-{\bf k^{''}})
W^*({\bf k'}-{\bf k^{''}})}  &
\end{eqnarray}
which encodes all the finite sampling hurdles for
recovering the power spectrum $C^l$, now to be examined. 

Firstly, the sampled modes power spectrum
$C^s({\bf k})={\langle a^s({\bf k})a^{s*}({\bf k})\rangle}$
becomes the convolution of the raw power spectrum
with the window power spectrum \cite{teggy,mike,peebles,gorski}. 
This has the effect
of leaving a low $k$ white noise tail in $C^s({\bf k})$ up to 
$k\approx 1/L$, and thereafter averaging out oscillations in the raw
spectrum on a scale $\Delta l\approx 1/L$. If the field has edges 
there will also be spurious oscillations of period
$1/L$ superposed on the spectrum. Field edges can be treated as in 
\cite{teggy}.
In the case of a square field the window should be  multiplied 
by a cosine bell.
Whenever the sampled spectrum is highly distorted,
a deconvolution recipe is then required.
In the presence of noise
and cosmic/sample variance this induces 
a large deterioration of the detection
function. We have however checked \cite{mikeandjoao} that, providing 
edge effects are treated, one has in the stCDM Doppler peak region
\begin{equation}\label{cscn}
C^s({\bf k})\approx C(k)B^2(k)
{\int {d{\bf k}'\over (2\pi)^2} |W({\bf k'})|^2 } = \alpha C(k)B^2(k)
\end{equation}
for fields with $L>4$ degrees, or $\theta_w>2$ degrees
(we illustrate this point in Fig.~\ref{fig1}).
Therefore, as long as the field is not too small, deconvolution
is trivial in the relevant spectrum sections of stCDM, 
and it does not add any extra errors. Note that $\alpha=\Omega^s
/(2\pi)^2$, where $\Omega^s$ is the field area 
($\Omega^s=L^2$ for a square
field,  $\Omega^s=\pi\theta_w^2/(8\log 2)$ for a primary beam).

Secondly, finite fields have the effect of correlating neighbouring 
$a^s({\bf k})$ modes within a correlation radius $\xi\approx 1/L$
(assuming edge effects have been treated)
or $\xi\approx1/\omega_w$. This
translates into a correlation length ${\overline\xi} (k)$
between $C(k)$ estimates. In \cite{mikeandjoao} 
we show how spectral resolution 
in the Doppler peak region is typically imposed, not by the fact that
a given $C(k)$ estimate receives contributions from neighbouring $k$,
but because we can only make uncorrelated estimates of the power
spectrum with a separation ${\overline \xi}(k)$. This effect is
reminiscent of cosmic covariance in non-Gaussian theories
\cite{ngsp}. Correlations also 
determine the cosmic/sample variance. Using 
${\rm cov}(x^2,y^2)=2{\rm cov}^2(x,y)$,
it can be proved that any power spectrum estimate $C_{\Omega}$ (using
a 2D region $\Omega$ of the Fourier domain, with area $A_k$,
in which $C^s({\bf k})$ does not vary
much) is affected by the sample variance
\begin{equation}\label{cv}
\sigma^2(C_{\Omega})={2C^2_{\Omega}\over {\tilde N}_{\Omega}}=
2C^2_{\Omega}{\int {d{\bf k}d{\bf k'}\over A_k^2}{\rm cor}^2
(a^s({\bf k}),a^{s*}({\bf k'}))}\; .
\end{equation}
${\tilde N}_{\Omega}$ acts as the effective 
number of independent modes in the region, and it can be used to define
an average density of independent modes $\rho_0$.

We have found it convenient to replace the ${\bf k}$ space by 
a square mesh, to be called uncorrelated-mesh, 
with a spacing locally given by  $k_0\approx 1/\sqrt{\rho_0}$. 
This mesh, on average, 
contains all the non-redundant information, given cosmic/sample
variance and the correlations
imposed by finite sampling. We have checked that
the uncorrelated-mesh is nearly a square lattice with 
$k_0\approx 2\pi/L$ for a square field, and $k_0\approx 2{\sqrt {4\pi\log 2}}
/\theta_w$  for a Gaussian field. Although it is easy
to improve on this approximation, it is normally a good enough recipe. 
Using only mesh points (denoted by ${\bf k}_i$)
the sampled power spectrum $C^s(k)$ can be estimated with
\begin{equation}
C_s(k)={1\over N_k}{\sum_{|{\bf k}_i|=k}}  |a^s({\bf k}_i)|^2
\end{equation}
where $N_k$ is the number of modes in the mesh which satisfy 
$|{\bf k}_i|=k$.
The residual correlations between these estimates fall below 5\%, but  
only a finite number of $k$ can be estimated.
Their average separation
$\Delta k\approx {\sqrt {k^2+k_0^2/\pi}}-k$, for $k>k_0$,
defines the maximal spectral resolution.
More estimates
could be inserted in between these, but they would necessarily 
be correlated estimates. Only for $k>k_0^2/(2\pi)$ 
can individual $C^l$ be estimated ($\Delta k\approx 1$).
The cosmic variance in these estimates
is approximately $\sigma^2(C_s(k))\approx 2C^{s2}(k)/N_k$.
For $k>k_0^2/(2\pi)$ this means 
$\sigma^2(C^l)\approx (C^{l2}/l)(4\pi/L^2)$, as naively expected
\cite{sample}.
For $k<k_0^2/(2\pi)$ the naive expectation breaks down.

We now study the effects of noise, differentiating between
single-dish experiments and interferometers. Let 
$\Omega_{pix}$ be the pixel area, and $\sigma^2_{pix}$
be the noise per pixel \cite{knox}. 
Fixing the detector sensitivity $s$
and total time of observation $t_{tot}$ fixes the quantity
$w^{-1}=4\pi s^2/t_{tot}=\sigma^2_{pix}\Omega_{pix}(4\pi/L^2)$,
which we therefore use to parameterize the noise level.
The noise introduces an extra diagonal term with value
$\alpha w^{-1} L^2/(4\pi)$ into the mesh modes covariance matrix.
Hence a centred uncorrelated-mesh estimator of the power spectrum is now
\begin{equation}
{\overline C}(k)={\left(
{1\over N_k \alpha B^2}{\sum_{|{\bf k}_i|=k}}  |a^s({\bf k}_i)|^2
\right)} -{\sigma^2_{pix}\Omega_{pix}\over B^2(k)}
\end{equation}
and its variance is
\begin{equation}\label{varnoise1}
{\sigma^2({\overline C}(k))\over C^2(k)}
={2\over N_k}{\left(1+
{w^{-1}L^2\over4\pi B^2C(k)}\right)}^2
\end{equation}
For interferometers \cite{tms} noise is added directly in Fourier space.
The noise in a given mesh cell is given 
by $\sigma^2_N= s^2\Omega^{s2}/(t_{vis} N_{vis})$,
where $N_{vis}$ is the number of visibilities in the cell,
$s$ is the sensitivity of the detectors, and 
$t_{vis}$ is the time spent observing each visibility. 
The coverage density $\rho_c=N_{vis}t_{vis}\Omega^s/t_f$ 
(where $t_f$ is the time
spent on a given field) is assumed to be uniform in a ring
of the $uv$-plane containing the stCDM relevant bins.
This assumes a dish geometry like the one proposed in \cite{intf}.  
If one decides to observe $n_f$ well-separated fields, then
each mesh-point acquires an extra index $i=1,\ldots, n_f$,
and points with different indices are uncorrelated. 
For fixed detector sensitivity and total observation
time one should now keep constant $w^{-1}=(2\pi)^2s^2/(\rho_c t_{tot})=
(2\pi)^2\sigma^2_N/(\Omega^{s3} n_f)$, 
and so this is the noise parameterization
we choose for interferometer estimates. 
A centred estimator is now
\begin{equation}
{\overline C}(k)={\left(
{1\over N_k \alpha }{\sum_{|{\bf k}_i|=k}}  |a^s({\bf k}_i)|^2
\right)} -{\sigma^2_N\over \alpha}
\end{equation}
and its variance is 
\begin{equation}\label{varnoise2}
{\sigma^2({\overline C}(k))\over C^2(k)}=
{2\over N_k}{\left(1+{w^{-1}\Omega^{s2} n_f\over C(k)}\right)}^2
\end{equation}
From these results one can compute $\Sigma$ in the large experiment
parameter space proposed (which always assumes $L>4^o$, or
$\theta_w >2^o$). Two types of results ensue.
Firstly, one can provide guidance on experimental design given a constraint,
such as finite funding. This constraint is mathematically
translated into hypersurfaces of constant $w^{-1}$.
Secondly, we may provide the value of the detection as a function
of $w^{-1}$, assuming ideal design. This will set lower bounds on
$w^{-1}$ for any meaningful detection, telling us thereafter how fast
the  detection improves with a given $w^{-1}$ improvement.


\begin{figure}[t]
\begin{center}
    \leavevmode
    {\hbox %
   {\epsfxsize = 8cm 
    \epsffile {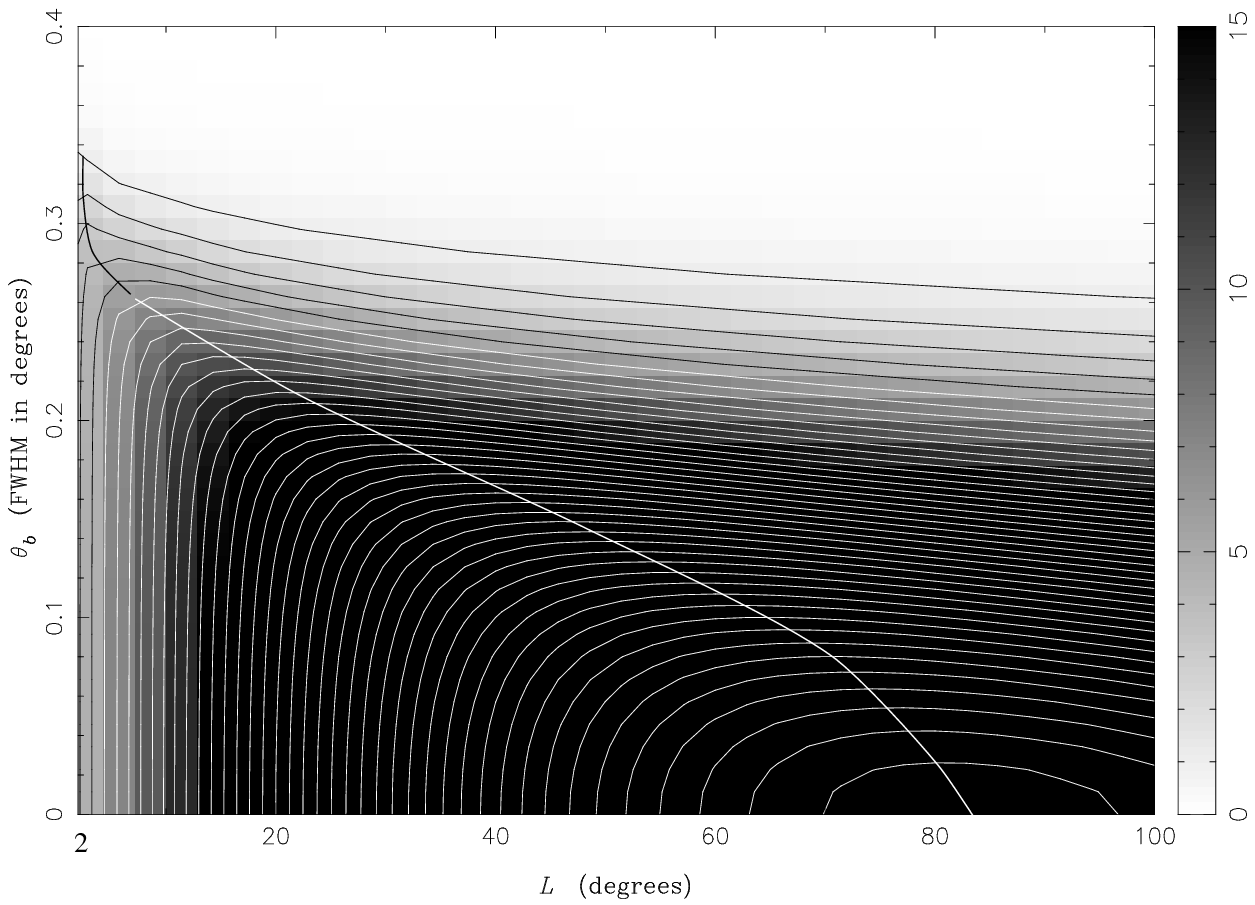} }}
	{\hbox %
   {\epsfxsize = 8cm 
    \epsffile {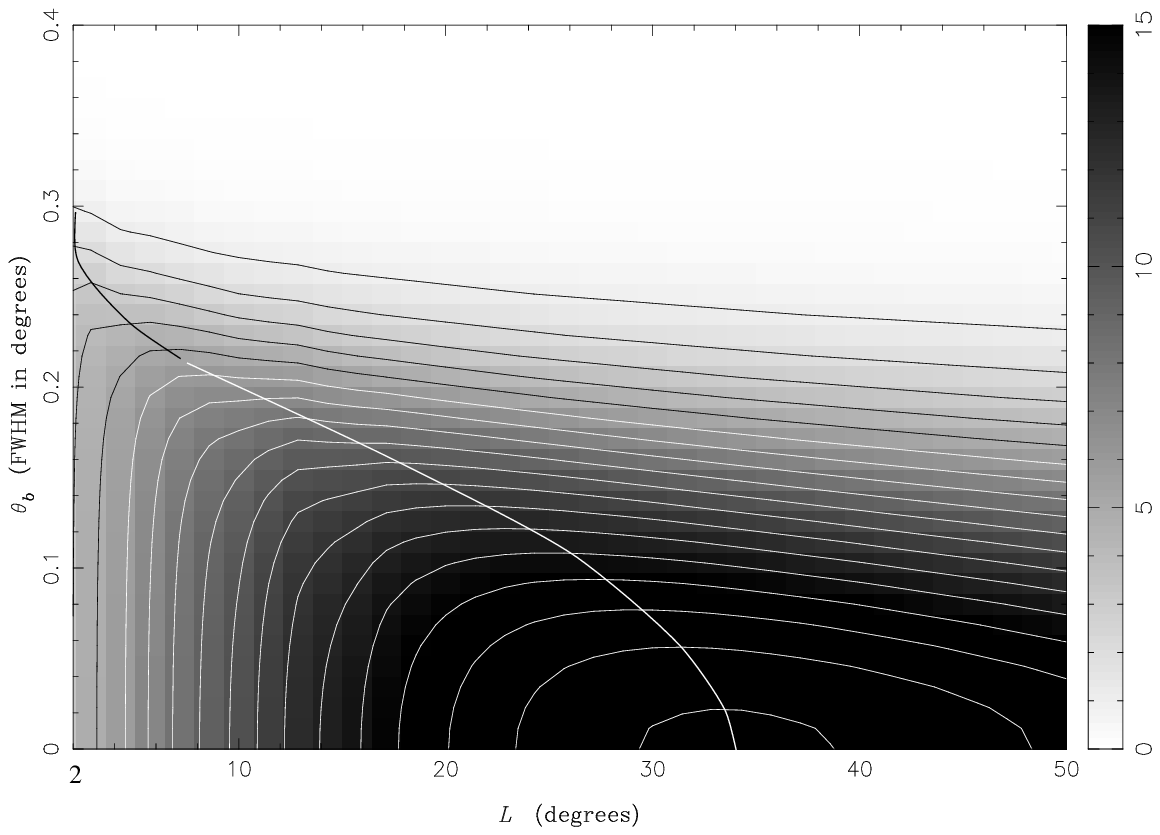} }}
\end{center}
\caption{Low noise ($w^{-1}=(25\mu K)^{2}(^o)^{2}$, top)
and high noise  ($w^{-1}=(60\mu K)^{2}(^o)^{2}$, bottom) contours
of the $\Sigma(L,\theta_b;w^{-1})$ function for stCDM. 
We have also plotted the ideal scanning lines $L_i(\theta_b)$.}
\label{fig2}
\end{figure}
In Fig.~\ref{fig2} we show a low and a high noise
section of $\Sigma(L,\theta_b;w^{-1})$ for single-dish experiments
\cite{comn}. 
Most noticeable are the high resolutions required for a significant
detection ($\theta_b<0.3^o$ and  $\theta_b<0.25^o$, respectively).
These are due to the fact that we are testing features at a rather large
$l$, and the noise term goes up exponentially with $l$ as we approach
the resolution limit.
It is also obvious that all-sky scanning is not ideal under realistic
levels of noise.
For fixed resolution and noise, increasing the coverage area
will at first increase the detection, but beyond a certain coverage $L_i$,
the detection will initially saturate, then start to decrease.
This is because noise separation relies on
allowing the {\it same}
coherent signal to compete with the incoherent noise. 
Only after the signal has dominated the noise does it make sense 
to increase the coverage area, so as to reduce the sample variance.
If the noise is very high, then all $t_{tot}$ should possibly
be used in a small patch of the sky (larger than $4^o$).
The ideal scanning lines $L_i(\theta_b)$  are plotted in Fig.~\ref{fig2}.
As the resolution increases so do $L_i$ and the 
achieved $\Sigma(L_i,\theta_b;w^{-1})$.
Initially they increase very fast; then, for $\theta_b<0.1^o$, not by much.
For a low noise  experiment ($w^{-1}=(25\mu K)^{2}(^o)^{2}$)
the detection increases from $\Sigma=1$ to $\Sigma\approx
36$ as the resolution is improved from $\theta_b=0.3^o$ to $\theta_b=0.1$
(with $L_i\approx 4^o$ and $L_i\approx 65^o$, respectively).
From then on $\Sigma$ improves by only a few sigmas. Even with infinite
resolution the ideal coverage area is $L\approx 80^o$. For 
a high noise experiment ($w^{-1}=(60\mu K)^{2}(^o)^{2}$)
the resolution is even more crucial. One needs $\theta_b=0.25^o$
to obtain a 3 sigma detection (with $L=4^o$), and  
a decent $\Sigma=10$ detection can be achieved
only with $\theta_b\approx 0.15^o$ (and $L_i=18^o$). 
It can be checked that all-sky coverage is only useful 
if the noise is lower than about $w^{-1}=(11\mu K)^{2}(^o)^{2}$.

\begin{figure}
\begin{center}
    \leavevmode
    {\hbox %
   {\epsfxsize = 8cm \epsfysize = 8cm
    \epsffile{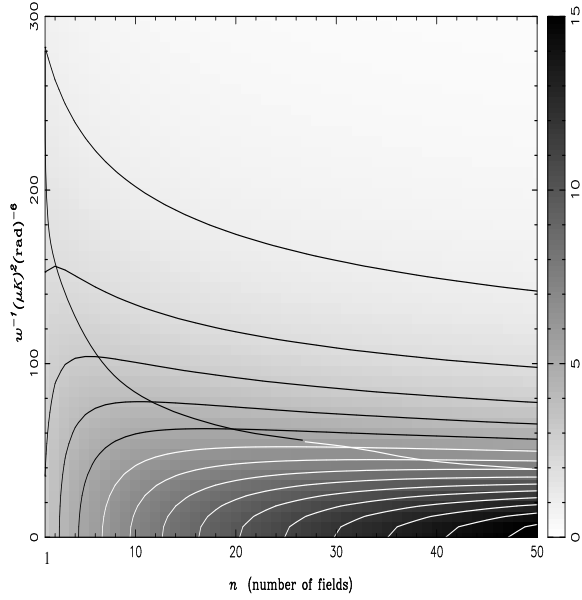} }}
\end{center}
\caption{$\Sigma(\theta_w=2^o,n_f;w^{-1})$ contours and density map,
and ideal scanning line $n_{fi}(w^{-1})$.}
\label{fig3}
\end{figure}

In Fig.~\ref{fig3} we show $\Sigma (\theta_w=2^o,n_f;w^{-1})$
for an interferometer. Ideal scanning now always means $\theta_w=2^o$,
and the ideal coverage area is fixed by a curve $n_{fi}(w^{-1})$. 
The high noise region is $w^{-1}>(150\mu K)^2{\rm rad}^{-6}$. There one 
should look into only one or two $2^o$ fields, in order to obtain 
a detection between 1 and 2 sigma. For $w^{-1}<(150\mu K)^2{\rm rad}^{-6}$ 
we enter the signal
dominated region. Following the ideal scanning line with decreasing
$w^{-1}$, $n_{fi}$ and $\Sigma$ start increasing, 
first slowly, then very fast.
For $w^{-1}=(100\mu K)^2{\rm rad}^{-6}$ one may obtain a 
3 sigma detection using 8 independent fields with $\theta_w=2^o$.
For $w^{-1}=(20\mu K)^2{\rm rad}^{-6}$
it is worth  looking into about 40 of these fields,
obtaining thus an 8 sigma detection. We have estimated CAT
noise level to be $w^{-1}=(300\mu K)^2{\rm rad}^{-6}$.
This is a mere prototype, and a 10-fold improvement should be easily
attained. 

These results stress the contradictions of an all-purpose experiment.
If the low-$l$ plateau of the spectrum is the theoretical target
then one needs all-sky coverage, and
satellite single-dish  experiments are to be favoured. 
Even if one wishes to target 
the main sCDM features, encoded mostly in the first peak's place
and height, then this is still true \cite{tegef}.
Our work shows how such a design relies heavily on the assumption that
the signal is in the vicinity of sCDM. If instead one is to test
the high-$l$ opposition between low $\Omega$ CDM and cosmic strings, 
then we have seen that single-dish experiments are
required to have rather high resolutions. 
Interferometers appear to be less constrained,
providing 2-3 sigma detections under very unassuming conditions,
with rapid improvements following further experimental condition
improvement.
Furthermore, in this context, all-sky scanning is not only
unnecessary, but in fact  undesirable. The best scanning
is normally achieved with deep small patches. These two features
contradict sharply the ideal experimental design motivated by the standard
theoretical gospel. We believe that a variety of contrasting experimental
techniques may equally well find their niche of important theoretical
implications.

We should mention, in closing, that if one is to combine the
high-$l$ cosmic string signal with the requirement that the low-$l$
section of the spectrum is to be measured, then the logic
is naturally changed. See \cite{andyben} for this alternative 
perspective.

ACKNOWLEDGEMENTS: J.M. would like to thank A.Albrecht, M.Jones,
and B.Wandelt for useful discussions. We thank M.White for
supplying the stCDM $C^l$ spectrum.
We acknowledge  St.John's College (J.M.),
and Trinity Hall (M.H.), Cambridge, for support in the form of 
research fellowships.

\end{document}